\documentclass[11pt,twoside]{article}

\usepackage{asp2006}
\usepackage{epsf}
\usepackage{psfig}
\usepackage{lscape}

\def\solm{M$_{\odot}\,$}

\def\solm{M$_{\odot}\,$}

\def\mass{$10^{11}$ M$_{\odot}\,$}
\def\hmass{$10^{11.5}$ M$_{\odot}\,$} 

\begin{document}
\title{The Assembly History of Massive Galaxies: What Do We Know?}   
\author{Christopher J. Conselice}   
\affil{University of Nottingham, England}    

\begin{abstract} 

Understanding the formation history of massive galaxies is one of most 
popular and longstanding problems in astronomy, with observations 
and theory 
addressing how and when these systems assembled.  Since the most massive
galaxies in today's universe, with M$_{*} > 10^{11}$ \solm,
are nearly all elliptical with uniform old stellar populations, 
we must probe higher redshifts to discover their full origins.  A recent
consensus has developed that nearly all M$_{*} > 10^{11}$ \solm
galaxies we see today were established by $z \sim 1$, with at most
a factor of two growth in stellar mass and number densities at lower
redshifts.  We review the evidence for this, and discuss how recent observations of 
star formation rates, colors, and morphologies of massive galaxies 
at $z < 1$ with M$_{*} >$ \mass show that these systems are still 
experiencing some evolution.  Massive galaxies undergo on average
a single major merger at $z < 1.5$, and roughly half are 
experiencing star formation at the same redshifts. The highest mass galaxies,
with M$_{*} >$ \hmass, appear in similar abundance at $z < 2$, suggesting
that extremely massive galaxies are mostly formed very early in the universe.  
Observations at $z > 1.5$ demonstrate
that major galaxy mergers are the primary method for assembling these 
massive galaxies, with nearly all of this merging occurring at $z > 2$,
with on average 4 to 5 major mergers taking place at $z = 1.5 - 3$.

\end{abstract}


\vspace{-0.5cm}
\section{Introduction}

Determining when and how galaxies in the universe formed is one of the 
most outstanding problems in cosmology and galaxy formation.   Galaxies 
are predicted in Cold Dark Matter based models of structure formation to 
assemble gradually with time through the merging of smaller systems (e.g,. 
White \& Rees 1978).  While there is some evidence for this 
process, at least in terms of galaxies (e.g., Le Fevre et al. 2000; 
Patton et al. 2002; Conselice et al. 2003a,b; Bridge et al. 2007), many
details are still lacking. Alternatively, massive galaxies, which
are mostly ellipticals in today's universe (e.g., Conselice 2006a), 
may have formed in a very rapid collapse of gas (e.g., Larson 1974).  

As such, massive galaxies are largely the test-bed for galaxy models.  
Understanding their evolution observationally is therefore an important 
test of the physics behind galaxy formation.  As star formation 
and merging activity has been seen in ellipticals from $z \sim 0$ to
$z \sim 1$ (Stanford et al. 2004; Lin et al. 2004; Teplitz et al. 2006), it
is not clear when or how the most massive galaxies finally assembled.
If it were possible to date every star in nearby massive galaxies, we 
could in principle determine the formation epoch and time-scales of these 
systems by examining their individual stars.  We cannot however resolve 
stars in all but the nearest galaxies, and their integrated stellar 
properties, such as  colors, become degenerate after about 5 Gyrs 
(e.g., Worthey 1994).   Stellar ages also do not necessarily correlate 
with the assembly of mass through, for example, merging activity (Conselice
2006b; De Lucia et al. 2006; Trujillo et al. 2006).  An alternative 
approach towards understanding massive galaxies and their evolution is 
empirically measuring the number densities, morphologies, star formation 
rates, and the stellar masses of massive systems at some fiducial 
time, and to compare these to similar quantities at different times 
(redshifts), and with models.

Observational evidence suggests that passively evolving massive galaxies 
exist at $z \sim 1$, and likely at even early times, at $z > 2$ (Fontana et al.
2004; Daddi et al. 2004; Glazebrook et al. 2004; Saracco et al.  2005).
Recent claims also exist for the establishment of the full massive galaxy
population
by $z \sim 1$ (e.g., Drory et al. 2005; Bundy et al. 2005, 2006; Borch 
et al. 2006; 
Cimatti et al. 2006). 
However, what is not yet clear is if number densities measured in these surveys
are able to rule out evolution at $z < 1$ due to uncertainties in 
measuring stellar masses, number densities, and cosmic variance. 

On the other hand, at $z > 1.5$
it appears that there are significantly fewer massive galaxies than
at $z < 1.5$ (e.g., Fontana et al. 2004).  Observationally, a large
fraction of the most massive galaxies at $z > 1.5$ are undergoing 
major mergers,
which are able to construct the stellar masses of these galaxies
rapidly (e.g., Conselice 2006b).   The merger history at $z < 1.5$ is not 
as clear, with observations
inconsistent on whether there is evolution in the massive galaxy population
at $z < 1$ (e.g., Bell et al. 2004; Brown et al. 2007; Scarlata et al.
2007).    We
argue in this review that after taking into account all sources of
error, and by examining the physical properties of massive galaxies
at $z < 1.5$, that at least one major merger is occurring within
these systems. A significant fraction of massive galaxies at $z < 1.4$ also
have not yet acquired a smooth elliptical structure, and have ongoing
star formation.  Thus, while the bulk of the stellar mass
in massive galaxies is present by $z \sim 1 - 1.5$, there is still 
observable evolution.  Throughout this review we use a standard cosmology of 
H$_{0} = 70$ km s$^{-1}$ Mpc$^{-1}$, $\Omega_{\rm m} = 
1 - \Omega_{\lambda} = 0.3$, and a Chabrier IMF for stellar mass
calculations.

\vspace{-0.52cm}

\section{Number and Mass Densities}

The most basic method for understanding the evolution of
massive galaxies is measuring how their number densities
and integrated mass densities change
as a function of time.   This is typically done through the use of
stellar masses (e.g., Brinchmann \& Ellis 2000; Conselice et al. 2005a,b;
Bundy et al. 2005,2006).  Alternatively, it has remained popular to
determine the number densities for luminous, red, or elliptical 
galaxies, although
these selection methods will produce biases when trying to understand the evolution of
massive galaxies, as star formation and morphological evolution are
occurring in ``early-type'' galaxies at $z \sim 1$ (e.g., Stanford et al. 2004;
Teplitz et al. 2006).  Recent work on measuring densities
suggests that within the uncertainties galaxies with large
stellar masses, with M$_{*} >$ \mass, are largely in place at $z \sim 1$ 
(Glazebrook et al. 2004; Bundy et al. 2005, 2006; Cimatti et al. 2006).
 
 \setcounter{figure}{0}
 \begin{figure}[!h]
 \plotone{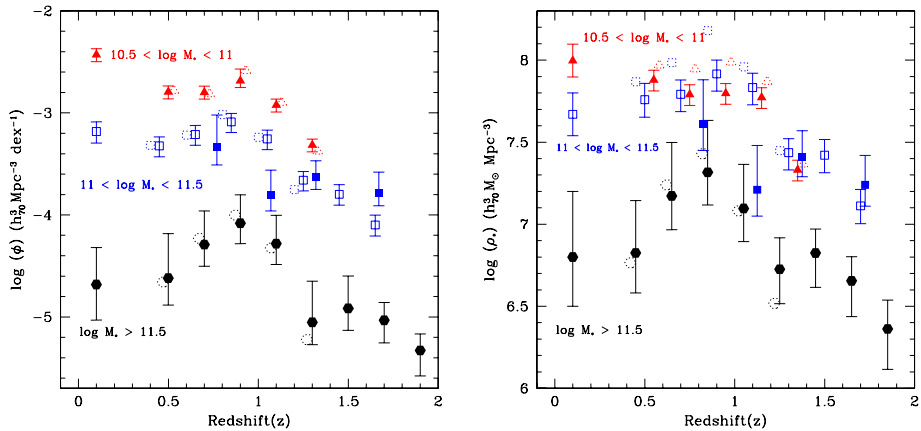}
 \caption{Left panel: the evolution in the number densities for galaxies of
various stellar masses between $z \sim 0.4 - 1.4$.  Right panel: the
stellar mass density evolution as a function of stellar mass at the same
redshift intervals.   The points at $z \sim 0$ are taken from Cole
et al. (2001).  The error bars listed on both the
numbers and stellar mass densities reflect uncertainties from stellar mass
errors, as well as cosmic variance, and counting statistics.  The
dashed symbols near each data point show how these values would
change if just using photometric redshifts.  For comparison we show the 
stellar mass densities
for systems with \mass$<$ M$_{*} <$ \hmass from Glazebrook et al.
(2004) plotted as solid blue boxes. Note that shifts of $\pm$0.05 in redshifts
have been applied so that the data points and error bars do not overlap.}
\end{figure}

Figure~1 shows an updated version of how the number and mass 
densities of galaxies with stellar
masses  M$_{*} > 10^{11.5}$ \solm and 
$10^{11}$~\solm~$<$~M$_{*}~<~10^{11.5}$ \solm evolve out to $z \sim 2$,
as seen in the large 1.5 deg$^{2}$ Palomar Observatory Wide-Field Infrared 
Survey (Conselice et al. 2007a) covering the DEEP2 fields 
(Davis et al. 2002, 2006) from 
Conselice et al. (2007b).     Figure~1 also shows the number densities of 
galaxies within these mass ranges measured
in the nearby universe to $z \sim 0.2$ by the 2MASS/2dF galaxy surveys 
(Cole et al. 2001), normalized using the same Chabrier IMF used for the
higher redshift stellar masses.  The number density evolution of these
massive galaxies demonstrates that statistically there is very little to no
evolution at $z < 1$ for the M$_{*} >$ \mass systems. This appears to support
the idea that massive galaxies are present by 
$z \sim 1$ (e.g., Cimatti et al. 2006; Bundy et al. 2006; 
Brown et al. 2007). 

However, as can be seen by eye in Figure~1, within the observational errors, 
there is some evolution in number densities for M$_{*} >$ \mass, and perhaps 
M$_{*} >$ \hmass selected galaxies between $z \sim 1-1.5$. The 
evolution in the number and mass densities can be examined quantitatively 
in a number of ways.  When considering evolution just within this
sample from $z = 1.5$ to $z = 1$ there are significant increases at 
masses \mass $< $ M$_{*} <$ \hmass, both in terms of number and mass densities. 
This is also the case when considering evolution between $z \sim 1.5$ 
and $z \sim 2$.  However, galaxies with M$_{*} >$ \hmass show an increase in 
number 
densities between $z = 1.5$ to 0.4 of a factor of 2.7$^{+1.8}_{-1.7}$. This is
significant only at the  $< 2 \sigma$ level, considering all 
uncertainties. In fact, all of this evolution occurs at $z > 1$.  
Furthermore, there is a factor of 1.3$^{+0.74}_{-0.53}$ increase in the mass 
density associated with  M$_{*} >$ \hmass galaxies at the same redshift range, 
although this is also at less than 3 $\sigma$ significance.  There is an 
increase of 11.2$^{+8.7}_{-4.9}$ in number densities, and
a factor of 5.5$^{+4.3}_{-1.7}$ increase in mass densities for systems
with M$_{*} >$ \hmass from $z \sim 2$ to $z \sim 1$. This is also an insignificant increase, and it is
impossible to rule out that massive galaxies with M$_{*} >$ \hmass are
all in place at $z < 2$.

An analysis of Figure~1 shows that the number densities
of systems with $10^{11}$ \solm $<$ M$_{*} < 10^{11.5}$ \solm increases
by a factor of 2.2$^{+0.57}_{-0.41}$ between $z = 1.4$ and $z = 0.4$, 
a result significant at $>$ 4 $\sigma$. Just as for the most massive
systems, this evolution occurs completely at $z > 1$.   Similarly, there
is a factor of 2.1$^{+0.6}_{-0.35}$ increase in the integrated mass density
for systems with $10^{11}$~\solm~$<$~M$_{*}~<~10^{11.5}$ \solm
within the same redshift range, also at $ > 4$ $\sigma$ confidence. After 
correcting for incompleteness there is a factor 
of 14.5$^{+4.1}_{-2.8}$ evolution in the
number densities, and a factor of 10.7$^{+3.1}_{2.0}$ in mass densities
between $z \sim 2$ and $z \sim 1$ for galaxies with  
$10^{11}$~\solm~$<$~M$_{*}~<~10^{11.5}$ \solm (Conselice et al. 2007b).

The observed evolution is such that the most massive systems with 
M$_{*} >$ \mass
increase in number and mass densities by factors $> 2-3$ at a significance
$> 3$ $\sigma$.  Taken as a whole, we calculate that the scenario whereby
the stellar mass and number densities of massive galaxies does not 
evolve between
$z \sim 1.5$ to $z \sim 0.4$ can be rejected at $> 8$ $\sigma$
confidence. Therefore it does not appear that high mass galaxy formation,
with the possible exception of M$_{*} >$ \hmass systems,
is complete by $z \sim 1.4$, yet it is largely completed by $z \sim 1$.   
Therefore,
the redshift range $z \sim 1 - 1.5$ is the final epoch for the build up
of the majority of the mass in massive galaxies.  However,
there could easily be a factor of 2 or 3 evolution in the mass and number
densities for massive galaxies at $z < 1$, and we would not be able to
measure this based on the current uncertainties in the measurements
of these quantities. The best way to approach this problem is to
determine if massive galaxies have any ongoing evolution based on
physical features, through structure and star formation.

\subsection{Structures and Morphologies}

Investigating the structures and morphologies of galaxies is becoming 
recognized as one of the most important methods for understanding 
galaxies (e.g., Conselice 2003; Cassata et al. 2005; Trujillo et al. 2006),
and for tracing the merger history at higher redshifts
(e.g., Conselice et al. 2003a; Bridge et al. 2007).

 \setcounter{figure}{1}
 \begin{figure}[!h]
 \plotone{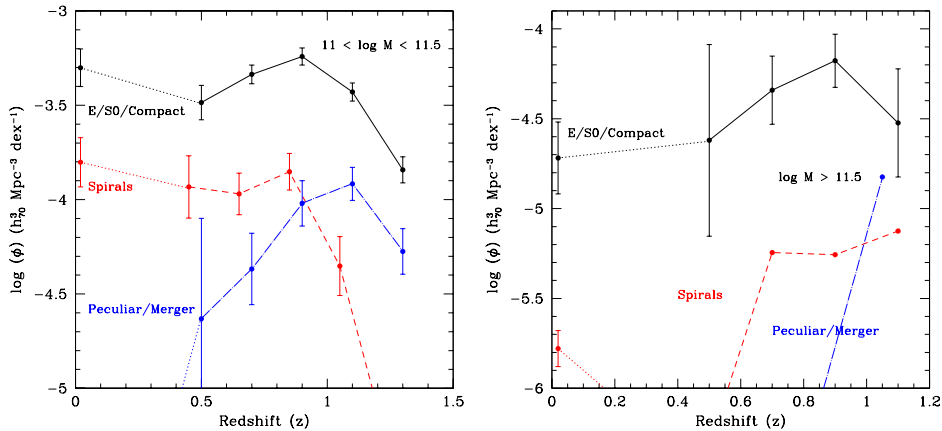}
 \caption{Visual estimates of the morphological evolution for
massive galaxies with M$_{*} >$ \mass taken from Conselice
et al. (2007b) using ACS data.  This figure is divided into two mass
ranges, and shows the evolution of galaxy number
densities at $z < 1.4$.  The left panels shows the morphological break down
up to $z \sim 1.4$
for systems with \mass $<$ M$_{*} <$ \hmass.  The right hand side shows the 
corresponding
evolution for galaxies with M$_{*} >$ \hmass.   The morphological
fractions and densities for the $z = 0$ systems are taken from
Conselice (2006a).}
 \end{figure}

Early work showed that massive galaxies at $z < 1$ are generally
early-types or disks (Brinchmann \& Ellis 2000; Bundy et al. 2005).
The overlap of Palomar 
NIR imaging and Hubble ACS imaging in the Extended Groth Strip (Davis
et al. 2006) allows us to study in detail $> 500$ galaxies with stellar 
masses M$_{*}$ $>$ \mass at $z < 1.4$.    For nearly all of
these systems, their magnitudes are bright enough such that effects
due to redshift do not affect the ability to classify these
systems either by eye, or through quantitative methods (e.g., Conselice
et al. 2000c; Windhorst et al. 2002; Papovich et al. 2005; 
Taylor-Mager et al. 2006; Conselice et al. 2007b).

Conselice et al. (2007b) find a significant amount of  morphological 
diversity among the M$_{*} >$ \mass galaxies (Figure~2). At $z < 1$, 
69\% of M$_{*}$ $>$ \mass systems are early-types (elliptical, S0, compact), 
while 10\% are disks, and 18\% are peculiars. This is perhaps a surprisingly 
high fraction of peculiars within a massive galaxy selected sample, and 
suggests that some of these systems are still undergoing some type of mass 
assembly, possibly through merging or star formation.  This changes slightly
when examining only the most massive systems with M$_{*} >$ \hmass.
These galaxies are $\sim 90$\% early-type over all redshifts, with a roughly 
similar number of mergers and disks making up the remainder.  These results
remain essentially the same, to within 5\%, after considering how the
Eddington bias may
bring lower mass galaxies into our mass cuts due to observational 
uncertainty (see also Brinchmann \& Ellis 2000; Bundy et al. 2005).

It is however clear that $\sim 30$\% of galaxies with M$_{*} >$ \mass
at $z < 1.5$ are not early-types, which suggests that there is evolution in the
massive galaxy population that cannot be seen simply through changes
in number and mass densities.  The disk galaxies show
that there is some star formation occurring, and the peculiars
reveal merger activity within this population.

Although early-types (classified E/S0/compact) dominate the massive
galaxy population at both the M$_{*} >$ \mass and  M$_{*} >$ \hmass  
selection limits, these galaxies often contain evidence for morphological 
peculiarities.   Usually these are in the form of outer low surface 
brightness features, or multiple cores.  A total of 68 out of 263 
(26$\pm3$\%) ellipticals with 
M$_{*} >$ \mass show some internal substructure visible by eye (Conselice
et al. 2007b).    These objects are perhaps seen in other ways, such as 
through color gradients and color structures in ellipticals (e.g., Menanteau
et al. 2005;  Stanford et al. 2004; Teplitz et al. 2006) resulting from 
star formation, and which may be related to these features. 
Previous studies have generally found that it is the lower mass ellipticals 
that contain these star formation signatures.  These
morphological disturbances however do not appear more common in the lower
mass ellipticals, and in fact, 36\% of the M$_{*} >$ \hmass ellipticals
show this signature - a higher fraction than in the M$_{*} >$ \mass 
population. These peculiarities are likely the result of recent merging
activity in these systems within the past 1-2 Gyr before we observe
them.


\subsection{Star Forming Properties of Massive Galaxies}

\underline{General Trends on the Color-Magnitude Diagram}: A major 
question concerning high-mass galaxies is whether
or not these systems have ongoing star formation at high
redshift. While it is commonly thought that massive and early-type
galaxies have largely finished their assembly and star formation by 
$z \sim 1$, detailed investigations suggest otherwise (e.g.,
Stanford et al. 2004; Teplitz et al. 2006; Conselice et al. 2007b).

One way to understand the star formation history of massive galaxies
is to examine their position on color-magnitude diagrams. 
We show in Figure~3  the M$_{\rm B}$ vs. $(U-B)_{0}$ 
diagram for M$_{*} >$ \mass galaxies at $z < 1.4$ taken from Conselice (2007b).   
Galaxies appear to separate into a red-sequence and a blue cloud 
in this parameter space (e.g., Strateva et al. 2001;  
Baldry et al. 2004; Bell et al. 2004; 
Faber et al. 2005).    At the highest redshift bin shown,
$1.2 < z < 1.4$, there is a significant number of
massive galaxies that are not on the red-sequence.  The fraction
of massive galaxies on the red-sequence however 
increases at lower redshifts.   This shows that massive systems 
with M$_{*} > 10^{11.5}$ \solm\,
generally fall in the red-sequence region at
all redshifts, but with a significant number of systems in the blue
cloud region at $z > 0.8$. The fraction of M$_{*} >$ \mass 
galaxies on the red-sequence increases with time at all
masses.  
Galaxies with lower masses show a similar pattern, yet lower mass galaxies
always have a lower fraction of galaxies on the red-sequence
at all redshifts, up to $z \sim 1.4$. 

 \setcounter{figure}{2}
 \begin{figure}[!h]
 \plotone{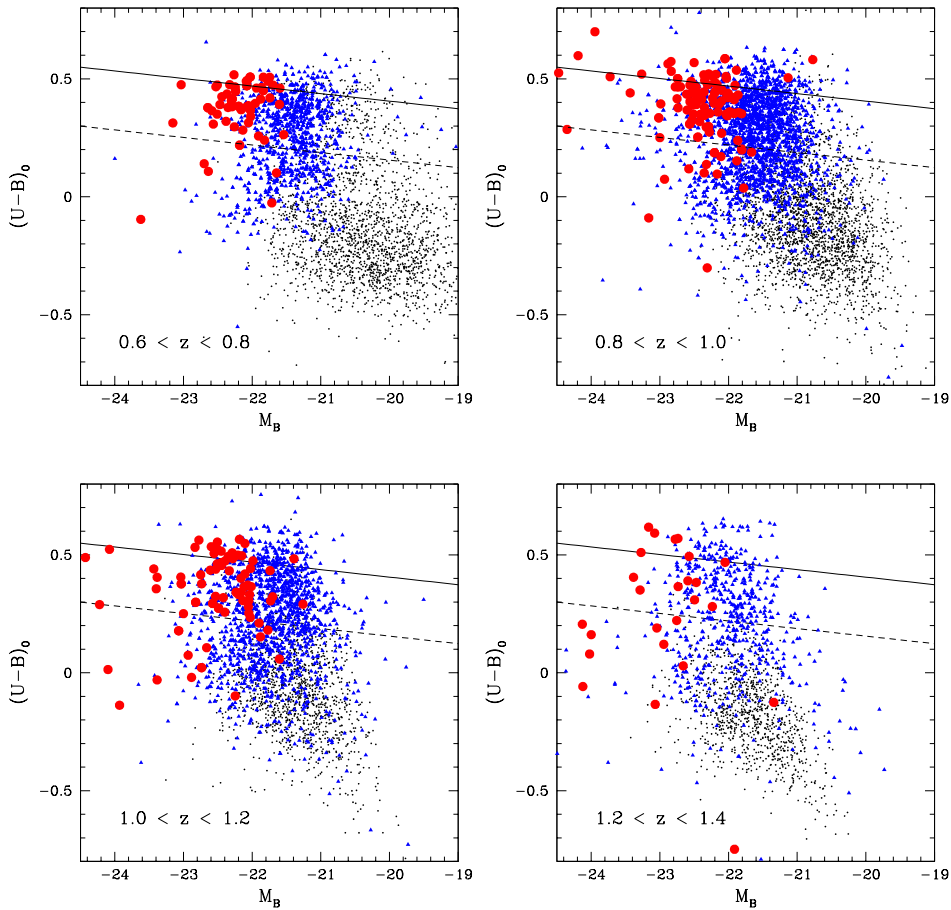}
 \caption{The $(U-B)_{0}$ vs. M$_{\rm B}$ diagram for galaxies with
M$_{*} > 10^{10.5}$~\solm from $z = 0.6$ to $z = 1.4$.  The
large red points on each panel are for galaxies with 
M$_{*} > 10^{11.5}$~\solm. The blue triangles show
the location of systems with $10^{11}$ \solm $<$ M$_{*} < 10^{11.5}$ \solm.
The solid line in each diagram
shows the location of the red-sequence, as defined in
Faber et al. (2005), and the dashed line is the demarcation
between red and blue galaxies.}
 \end{figure}

This leads to a very important conclusion regarding the
red-sequence and high mass galaxies. Previous studies have
examined the increase of the amount of stellar mass on the red-sequence,
finding as much as a factor of two increase 
since $z \sim 1$ (Bell et al. 2004; Faber et al. 2005; Brown et al. 2007). 
However, this increase is due to galaxies appearing on the red-sequence,
which were previously blue, and not due to in-situ growth on
the red-sequence itself. This can be clearly seen by massive galaxies  
gradually moving onto the red-sequence with
time.  This effect is also revealed in the decline in the number of
blue massive galaxies found in the universe since $z \sim 1$ (e.g.,
Bundy et al. 2006).  This is not consistent with the idea that
the red-sequence grows solely through the so-called `dry mergers'.
Although merging may be present within the red-sequence, and within our
massive galaxy sample, it does not appear to be the dominate
method whereby the red-sequence grows.  

\vspace{0.25cm}

\noindent \underline{Star Formation Rates}: Quantifying star formation in 
massive galaxy samples can be done in 
several ways, including rest-frame UV emission, emission line fluxes,
and Spitzer MIPS 24 $\mu$m data.  Perhaps surprisingly, about half of all
massive galaxies at $z \sim 1$ are detected at 24 $\mu$m, after removing
AGN contamination based on Chandra detections (Conselice et al. 2007b).
After matching the MIPS and [OII] star formation indicators,
Conselice et al. (2007b) find that $\sim$40\% of the M$_{*} >$ \mass 
systems at $0.4 < z < 1.4$ are detected at 24 $\mu$m.  A total of 
37$\pm5$\% of the systems with M$_{*} >$ \hmass within this redshift
range are detected
at 24 $\mu$m, with an average star formation rate of 70 \solm yr$^{-1}$.

For galaxies with stellar masses \mass $<$ M$_{*}$ $<$ \hmass,
Conselice et al. (2007b) find that the fraction of galaxies undergoing 
star formation remains roughly similar at all redshifts. Circumstantially, 
this is consistent with the fact that the fraction of spirals+peculiars
in this mass cut remains roughly constant throughout this redshift
range. Interestingly, the fraction of systems which are undergoing 
star formation is higher than the non-elliptical fraction, showing that
some morphologically classified massive ellipticals must be undergoing star
formation (Stanford et al. 2004; Teplitz et al. 2006).  The fraction of 
M$_{*} >$ \hmass galaxies with a significant 24 $\mu$m detection declines 
slightly at lower redshift,  from 33\% at $1.2 < z < 1.4$ to 
14\% at $0.4 < z < 0.6$, consistent with a drop in the 
morphological fraction of non-ellipticals. This is however certainly a 
lower limit to the evolution, as the number
of galaxies detectable at 24 $\mu$m declines at higher redshifts. 

Conselice et al. (2007b) find, similar to previous studies
utilising IR star formation indicators (e.g., Le Floc'h
et al. 2005), a decline with redshift for the massive galaxy population.
After fitting these star formation histories up to their
plateau (i.e., at $z \sim 1$) as a power-law $\sim (1+z)^{\alpha}$
we can quantify the star formation history differences 
between M$_{*} >$ \mass galaxies  and M$_{*} >$ \hmass
galaxies.  For systems with M$_{*} >$ \hmass, Conselice et al. (2007b)
find that the star formation rate declines as $\alpha = 6\pm2.2$,
and for systems with \mass $<$ M$_{*}$ $<$ \hmass the slope is fit as
$\alpha = 4.1\pm0.64$.  The overall decline in the
entire galaxy population's star formation history can be
parameterised as $\alpha = 3-4$ (Hopkins 2004; Le Floc'h et al. 2005). 
It appears that while
the \mass $<$ M$_{*}$ $<$ \hmass galaxies have a similar
decline as the overall field, the highest mass galaxies
show a faster decline.

\vspace{-0.5cm}

\section{Galaxy Merging}

A major question concerning massive galaxies is the role of mergers
in their formation. Galaxy mergers are occurring in the universe, and
it is likely that they play some role in the formation of massive
galaxies, but the details are still debated. If
mergers are occurring at $z < 1$, there are very few of them, and they
might be nearly all dissipationless 'dry' mergers, without star 
formation.  On the other hand, major mergers appear to be the dominate
method for forming massive galaxies at $z > 2$.

\vspace{0.25cm}

\noindent \underline{CAS Structural Analysis and the Merger Rate}: The 
CAS (concentration, asymmetry, clumpiness) parameters allow us to
probe the structures of galaxies quantitatively, and are a major method
for determining mergers in a galaxy population (e.g.,
Conselice et al. 2000a,b; Bershady et al. 2000; Conselice et al. 2004, 2005; 
Conselice 2003; Casatta et al. 2005; Bridge et al. 2007). The CAS system 
can also be used to identify relaxed massive
ellipticals.  The basic idea is that galaxies have light distributions
that reveal their past and present formation modes (Conselice 2003). 
One benefit of using the CAS system for
finding mergers is that it allows us to quantify the
merger rate and the number of mergers occurring in
a galaxy population (Conselice et al. 2003a; Conselice 2006b).

The location of massive galaxies with M$_{*} >$ \mass at $z < 1.4$ 
are generally found in CAS
space at the locations where they are expected based on their
visual morphologies.  One important exception is that many of the 
visually classified non-distorted early-types are not located in the 
corresponding $z \sim 0$ part of CAS diagrams, being slightly
too asymmetric (Conselice et al. 2007b). 

Using CAS values for massive galaxies with M$_{*} >$ \mass Conselice
et al. (2007b) determine the evolution of the non-dry merger fraction for 
massive galaxies (cf. Hernandez-Toledo et al. 2006 for understanding
dry mergers within CAS).  By using the the criteria, outlined in 
Conselice (2006b) of,

\begin{equation}
{\rm A > 0.35,\, A > S.}
\end{equation}

\noindent Conselice et al. (2007b) determined the merger fraction 
for the M$_{*} >$ \mass galaxies out to $z \sim 1.4$.  There is a 
slight decrease with redshift in the 
merger fraction such that it declines as $(1+z)^{1.3}$, similar
to the evolution seen in lower mass galaxies at $z < 1$ 
(Conselice et al. 2003;  Bridge et al. 2007).

Using the number densities of massive systems, and
time-scales for CAS mergers, we can calculate
the merger rate for M$_{*} >$ \mass galaxies based on the merger fraction,
and the time-scales for merging derived from N-body models.  From
this, a major merger time-scale of $\tau_{\rm} = 0.43\pm0.05$ Gyrs for a 
galaxy with a mass of 3$\times 10^{11}$ \solm is calculated (Conselice
et al. 2007b).

The merger rate of massive galaxies at $z < 1$ can then be calculated through
the merger rate equation,

\begin{equation}
{\rm \Re(z) = f_{m}(z) \cdot \tau_{m}^{-1} n_{m}(z)}
\end{equation}

\noindent where n$_{\rm m}(z)$ is the number densities of
objects, and f$_{\rm m}(z)$ is the merger fraction\footnote{Note that
this is not the galaxy merger fraction, which is
the fraction of galaxies merging, which is roughly 
double the merger fraction (Conselice 2006b).}  We
find that, statistically, the merger rate for these
M$_{*} >$ \mass galaxies is constant from 
$z \sim 0.4 - 1.4$, and is on average
log $<\Re>$ (Gyr$^{-1}$ Gpc$^{-3}$) = $4.3^{+0.4}_{-0.7}$.

We can furthermore calculate the total number of major
mergers a galaxy with M$_{*} >$ \mass undergoes
from $z \sim 1.4$ to $z \sim 0.4$ using equation (11)
in Conselice (2006b).  We calculate that the average
number of mergers a massive galaxy with M$_{*} >$ \mass
undergoes from $z \sim 1.4$ to 0.4 is N$_{\rm m} = 
0.9^{+0.7}_{-0.5}$.  Thus, on average, a massive
galaxy will experience about one major merger from $z \sim 1.4$
to $0.4$, roughly consistent with other results
(Conselice 2006b; Bell et al. 2006).

\vspace{0.25cm}

\noindent \underline{Dry-Merging at $z < 1$}: One issue which is not clear is how much of the merger and star formation
process, and especially the controversial and hard to find dry mergers, are 
responsible for the addition of mass in massive galaxies at $z < 1$.  
We can address this using mass functions, and the measured 
star forming histories of galaxies with M$_{*} >$ \mass and 
\mass~$<$~M$_{*}~<$~\hmass.  While the star formation history matches the
increase of the stellar mass, within massive galaxies, to within 
$< 3~\sigma$ at any
one redshift, the fact that the star formation history is consistently 
lower implies that star formation statistically cannot 
account for the total increase in stellar mass.   This implies that part of 
the mass growth in these systems must be accounted for by mergers, or 
galaxies with masses lower than each stellar mass
limit evolving into the higher mass bin due to star formation and/or
merging.
 
Between the bins M$_{*} >$ \hmass and  \mass~$<$~M$_{*}~<$~\hmass
the amount of stellar mass added to the higher mass bin can 
be measured partially through the star formation rate. The star
formation rate within a bin will increase the amount of mass
within that bin, and star formation in a lower mass bin
will increase the mass and number densities in higher mass
bins by bringing up galaxies.  When comparing
the changes in the mass function to the amount of new mass from star
formation, it is clear that changes in mass and number densities cannot 
be totally
accounted for by just star formation.   The remainder of the excess
must be produced through merging.  In Conselice et al. (2007b) it is
argued that the amount of merging is such that at least one major
merger is occurring for M$_{*} >$ \hmass galaxies at $z < 1.5$. Since
these galaxies are largely early-types, it is likely that many of these
mergers are dry (see also Bell et al. 2005).  This roughly agrees
with the number of mergers calculated through the CAS parameters.

\vspace{0.25cm}

\noindent \underline{High-redshift $z > 2$ mergers}: The situation at higher 
redshifts however appears 
to be different, and it is likely that the majority of the stellar mass 
in  modern ellipticals was put into place through major mergers
at $z > 1.5$.  Massive galaxies at $z > 2$ are not smooth ellipticals,
but appear peculiar, even in the rest-frame optical (Conselice et al.
2005).  In fact, the CAS values for these galaxies reveals a merger
fraction of roughly 40-50\% (Conselice et al. 2003).

 \setcounter{figure}{3}
 \begin{figure}[!h]
 \plotone{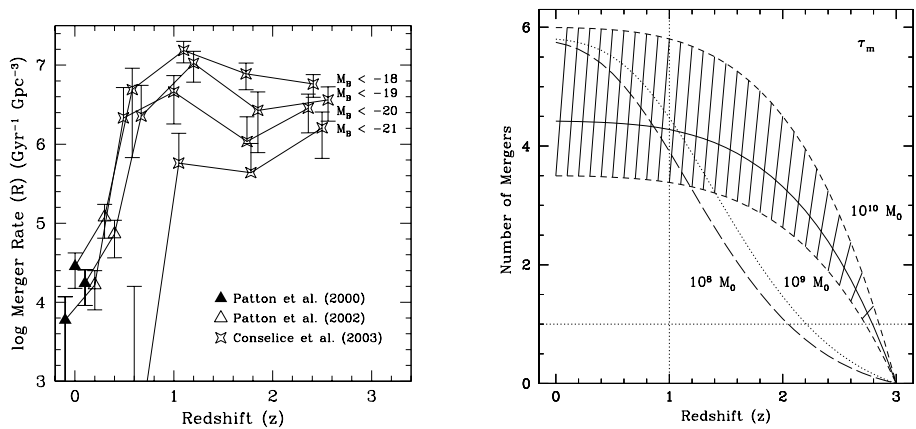}
 \caption{Evolution of the merger rate, in units of Gyr and co-moving
Gpc$^{3}$,
as a function of redshift and observed magnitude (left panel), and the 
empirically determined integrated number of major mergers since 
$z \sim 3$.  These merger rates and histories are taken from
previously published merger fractions (see Conselice 2006b and references
therein.)}
 \end{figure}

By integrating the merger rate since $z \sim 3$ Conselice (2006b)
finds that a typical massive galaxy with M$_{*} > 10^{10}$ \solm undergoes 
4.4$_{-0.9}^{+1.6}$ mergers from $z = 3$ to $z = 0$ (Figure~4). Most
of these mergers are at $z > 1.5$. An additional feature of 
the N-body models 
analyzed in Conselice (2006b) is the ability to determine the merging
galaxy mass ratios that can produce high asymmetries.  
The result of this is that the CAS method is only
sensitive to major mergers, that is mergers with a mass
ratio of 1:3 or lower (see also Hernandez-Toledo
et al. 2005).   This also allows us to determine how much
mass is likely added to galaxies due to the merger process
since $z \sim 3$. The result is that a galaxy which undergoes
on average 4 - 5 major mergers will increase its total mass by
a factor of $\sim10$. This is consistent with direct observations
which show that the most massive galaxies are generally in place
by $z \sim 1.5 - 2$, but are significantly depleted in number
at $z > 2$ (Fontana et al. 2004).

\vspace{-0.3cm}

\section{Discussion and Summary}

The last few years have seen a number of studies designed to determine when
massive galaxies in the universe formed.  These studies have generally 
concluded that massive galaxies in the universe, typically those with 
M$_{*} >$ \mass, are largely formed by $z \sim 1$, but with considerable 
uncertainty.   This is statistically
found to be the case in nearly all studies in terms of mass and number 
densities, for systems 
with M$_{*} >$ \mass from $z \sim 1$, although there is measurable 
evolution from $z \sim 1.4$ (Conselice et al. 2007b).

While studies have found that the number and mass densities of
massive galaxies are similar at $z < 1$, this does not necessarily imply
that there is no evolution in this population.  Using results from the wide 
and deep Palomar Observatory Wide-Field 
Infrared Survey, combined with DEEP2 spectroscopy, we can directly
select and study the properties and evolution of M$_{*} >$ \mass galaxies 
at $0.4 < z < 1.4$.  Based on the findings of this survey (Bundy et al.
2006; Conselice et al.
2007b) it appears that the stellar mass and  number densities of 
M$_{*} >$ \mass galaxies does not change significantly at $z < 1$.  
We however cannot rule out factors of 2-3 in number and mass density 
evolution for these systems, based solely on densities, due to the 
uncertainties in these measurements.  

Other methods besides densities, are therefore need to conclusively argue 
whether massive galaxies are finished forming by $z \sim 1$.   The fact 
that a high fraction of
massive galaxies are forming stars (40\%), and are non-elliptical
(30\%), at $z \sim 1$ suggests that there is active evolution.  We find that in
addition to star formation activity, a typical massive galaxies with
M$_{*} >$ \mass will undergo, on average, a single merger at $z < 1.4$.  Most
of the formation for these massive galaxies occurs at higher redshifts.
Observationally, there is a significant decrease in the number densities
of massive galaxies at $z > 2$. These systems are also observed to
be undergoing a significant amount of merging which is likely how they 
build up most of their mass by $z \sim 2$.
  
Furthermore, these results show that the study of `early-types', defined
through luminosity, color, morphology or stellar mass, at
high redshift must be carefully done, and results of studies
will vary significantly, depending on selection.  It is
clear, particularly at high redshift, that red galaxies are
not the equivalent of massive galaxies, or elliptical galaxies,
and each of these populations must be studied individually.  

\acknowledgements 

I thank the members of the Palomar, DEEP2, AEGIS, and Spitzer teams
for their contribution to this work. Support from a NSF Astronomy
and Astrophysics Fellowship and PPARC are gratefully acknowledged.


\end{document}